\documentclass[a4paper,12pt]{article}
\usepackage[utf8]{inputenc}
\usepackage[english]{babel}
\usepackage{graphicx}
\usepackage{amsmath}
\usepackage{relsize}
\usepackage[singlespacing]{setspace}
\usepackage[headheight=1in,margin=1in]{geometry}
\usepackage{fancyhdr}
\usepackage{float}
\usepackage{wrapfig,booktabs}
\usepackage{parskip}
\usepackage[colorlinks=true, allcolors=blue]{hyperref}
\usepackage{epstopdf}
\usepackage[colorinlistoftodos]{todonotes}
\usepackage{comment}
\usepackage{authblk}

\title{Towards Entrepreneurial Ecosystem Indicators: \\ Speed and Acceleration}
\author[1]{Th\'eophile Carniel}
\author[1,2]{Jean-Michel Dalle}

\affil[1]{Agoranov, Paris, France}
\affil[2]{Sorbonne Universit\'e, Paris, France}

\date{}

\begin{document}

\maketitle

\begin{abstract}
    We suggest the use of indicators to analyze entrepreneurial ecosystems, in a way similar to ecological indicators: simple, measurable, and actionable characteristics, used to convey relevant information to stakeholders and policymakers. We define 3 possible such indicators: \textit{Fundraising Speed}, \textit{Acceleration} and \textit{n\textsuperscript{th}-year speed}, all related to the ability of startups to develop more or less rapidly in a given ecosystem. Results based on these 3 indicators for 6 prominent ecosystems (Berlin, Israel, London, New York, Paris, Silicon Valley) exhibit markedly different situations and trajectories. Altogether, they contribute to confirm that such indicators can help shed new and interesting light on entrepreneurial ecosystems, to the benefit of potentially more grounded policy decisions, and all the more so in otherwise blurred and somewhat cacophonic environments.
\end{abstract}

\section{Introduction}
\label{sec:introduction}
In ecology, \emph{ecosystem indicators} are quantitative measurements that serve as proxies for some of the major ecological characteristics of an ecosystem, when these characteristics are of scientific and/or management importance but are challenging to measure directly~\cite{EPA2008, Kurtz2001, Landres1988, Fleishman2009}. For instance, the number of different taxa of a given species can be used as an indicator of biodiversity in a given ecosystem; populations of some species, such as lichens or amphibia, can serve as an indicator of pollution; etc. Ecosystems are intrinsically complex: indicators help describe them in measurements that can be understood and 'actionably' used by non-scientists to make well-informed management decisions.


\emph{Entrepreneurial ecosystems} have become a common notion to describe the complex geographical concentration, and interdependence, of the many different actors necessary to the creation and development of innovative startups~\cite{Autio2014}. This notion was initially embraced by businesses and public decisionmakers, in order to try to improve the conditions under which entrepreneurship and innovation can "thrive", i.e. to foster more "vibrant" ecosystems. Entrepreneurial ecosystems have rapidly become central in local, national and international innovation policies~\cite{Gilbert, Minniti}, as innovative startups were drawing ever-increasing investments from venture capitalists and benefited from an increased attention because of their potential to create jobs and economic growth~\cite{Birch,Thurik,Wong}. Following the leading example of San Francisco and the Silicon Valley, Austin, Boston, Los Angeles or New York in the US, Berlin, London or Paris in Europe~\cite{Slush}, among numerous others, have started competing against one another as entrepreneurial ecosystems, in order to support startups and to attract investors.

Much about the notion of entrepreneurial ecosystem is obviously yet problematic, and economics and management sciences have seriously started to analyze it both theoretically and empirically (e.g. \cite{Audretsch2018}). Rapid progress in both of these dimensions is much needed in order not only to assess the validity of this notion, but also, most notably, to provide a clearer understanding of the conditions under which entrepreneurial ecosystems do "thrive" or become "vibrant". Until then, we are left with the situation where policymakers have to make policy decisions with respect to entrepreneur ecosystems based only on scant and anecdotal information they are able to gather with respect to their and other ecosystems. Most of this information is unfortunately  vague and imprecise, and can be biased since it often originates from ecosystem actors themselves, whose incentives are thus strategically biased with respect to policymakers.

Furthermore, this difficult situation is reinforced by the fact that many notions circulate widely in the media and in the jargon of entrepreneurial ecosystems whose grounding appears weak and probably disputable. Most prominent among these weakly grounded notions often referred to with respect to entrepreneurial ecosystems, is the notion of \textit{acceleration}. Startups and ecosystems "accelerate": they develop more quickly, whatever quickly might mean, and without any quantitative assessment. This notion of acceleration has been specially promoted by so-called accelerators, a new kind of early-stage investors who offer a selective program targeting nascent startups, but no consensus has yet emerged about what acceleration actually means, and whether it is simply achievable beyond the promise made by its most vocal proponents.

It is in this context that we would like to suggest that ecosystems indicators, as used in ecology, could be of special interest with respect to entrepreneurial ecosystems. Until a better and deeper theoretical and empirical understanding of entrepreneurial ecosystems is reached, rigorously grounded indicators could help policymakers determine policies in a better-informed and actionable way, and measure their impact. We basically understand here \textit{ecosystem indicators}~\cite{Wicks} as simple, meaningful and communicable quantitative metrics that measure relevant characteristics of an ecosystem, in order to communicate with decision-makers in a more efficient and objective manner, and to help them determine more appropriate courses of action.

Based on the fundraising speed of startups and using data from \hyperlink{https://www.crunchbase.com}{Crunchbase}~\cite{dalle2017}, we notably propose an empirically-grounded, quantitative measurement of \textit{acceleration}, as a potential indicator for entrepreneurial ecosystems. We thus compare 6 leading entrepreneurial ecosystems and show that an approach based on indicators can relatively easily help understand important aspects of their dynamics, through time, and can potentially shed more light on related policies and acceleration strategies.

In order to do so, Section 2 defines indicators, notably \textit{fundraising speed} and \textit{acceleration}; Section 3 presents related measurements for 6 prominent entrepreneurial ecosystems in the world, before Section 4 provides an interpretation and concludes.




\section{Indicators}
\label{sec:indicators}

\subsection{\textit{Fundraising Speed}}
\label{subsec:speed}
Raising funds is crucial for startup development. The speed at which a startup can hire new employees, develop its operations, its sales, its R\&D, crucially depends on the pace at which it is able to raise funds.
Raising capital quickly is key to the growth of startup companies.
The ability to raise funds is of course influenced by characteristics of any given startup, notably by the experience of its founders, by its competition, etc. But, independently of these idiosyncratic criteria, the speed at which startups raise funds \textit{within a given ecosystem} is a key characteristic of this ecosystem, taken into account by entrepreneurs and founders when assessing the dynamism and vibrancy of any ecosystem, and a key element that policymakers try to evaluate and to influence so as to foster faster fundraising and hence faster development for the startups in the ecosystem they are in charge of.

By analogy with physics, we simply define the \textit{fundraising speed} of a startup as the difference between two measures of funding, averaged over the time elapsed between the two measures. Here, the \textit{fundraising speed} $v(i,t)$ of startup $i$ at time $t$ is defined by :

\begin{equation}
\mathlarger{
v(i, t) = \frac{\Delta f(i,t)}{\Delta t} = \frac{\mathlarger{\sum\limits_{t' = t_{0}}^{t}} f(i, t') - f(i, t_{0})}{t - t_{0}}
}
\end{equation}

where $f(i, t)$ is the total amount of funding raised by startup $i$ at time $t$ after its creation: $t_{0}$ is the creation date of the startup. In this paper, we standardize the measurement of $t$ in days and of $f$ in US dollars for all startups in all ecosystems.

We then define the \textit{fundraising speed} $V_{e}(t)$ of ecosystem $e$ at time $t$, understood as a first indicator of the dynamism and vibrancy of an ecosystem, as the median\footnote{We use median values in this paper rather than means due to the non-gaussian nature of some of the underlying distributions. In addition, it should also be noted that similar results to the ones presented here hold for measures based on the first decile of the distributions, instead of their median.} of all speeds $v(i,t)$ of all startups i in the ecosystem at time $t$, i.e. the median speed of the startups in this ecosystem $t$ years after their creation:

\begin{equation}
\mathlarger{
V_{e}(t) = Median(v(i,t))\ \forall i \in e
}
\end{equation}

\subsection{\textit{Acceleration} and \textit{n\textsuperscript{th}-year Speed}}
\label{subsec:acceleration}
We next define the \textit{acceleration} of a startup as the difference of fundraising speed between two measures of fundraising speed, averaged over the time elapsed between these two measures. The acceleration $a(i, t)$ of startup $i$ at time $t$ is thus given by:

\begin{equation}
    \mathlarger{
    a(i, t) = \frac{\Delta v(i,t)}{\Delta t}
    }
\end{equation}

Accordingly, we define the \textit{acceleration} $A_{e}(t)$ of ecosystem $e$ at time $t$, understood as a second indicator of its dynamism and vibrancy, as the median of the accelerations of all startups $i$ in ecosystem $e$ at time $t$.

\begin{equation}
    \mathlarger{
    A_{e}(t) = Median(a(i,t))\ \forall i \in e
    }
\end{equation}

We further compute the \textit{n\textsuperscript{th}-year speed} for year $y$ of ecosystem $e$, denoted as $v_n(e,y)$, defined as the median fundraising speed of startups founded in year $y$, measured during year $y+n$, i.e. the median fundraising speed of startups founded during year $y$, $n$ years after their creation at time $t_{0}$:

\begin{equation}
    \mathlarger{
    v_n(e,y) = Median(v(i,n))\ \forall i \in e \mid t_{0}(i) = y
    }
\end{equation}

For each ecosystem, this third indicator offers a complementary approach to acceleration by allowing to observe variations of $v_n(e, y)$ across year spans. Suppose that we compute $v_1(New York,y)$ for each $y \in [2010, 2015]$. We can then compare the median fundraising speed of startups founded in 2010 \textit{one year after their creation} ($v_1(New York,2010)$) with the median fundraising speed of startups founded in later years \textit{one year after their creation} (for instance, $v_1(New York,2014)$). Thus, we can determine whether the first-year fundraising speed of startups in New York increases over time, i.e. whether the New York ecosystem \textit{accelerates} over time.



\section{Results}
\label{sec:results}

\subsection{\textit{Fundraising Speed}}
\label{subsec:comparative_results}
We measure \textit{Fundraising Speed} as defined in section~\ref{sec:indicators} for 6 prominent worldwide ecosystems -- Berlin, Israel, London, New York, Paris (Ile-de-France Region), Silicon Valley --, using a dataset extracted through the Crunchbase API on January 6th, 2020. This dataset contains information on startups (name, creation date, headquarter location, sectors of activity), funding events (target startup, date, investors involved, amount, investment stage), investors (name, creation date, investor type, investor location) and individuals (name, past and current professional experiences, level and sectors of education, company board memberships and advisory roles). We thus have access to data on 472 startups for Berlin, 789 for Israel, 2081 for London, 2815 for New York, 874 for Paris and 6094 for the Silicon Valley.


More specifically, we:
\begin{itemize}
    \item measure the speed $v(i,t)$ of each startup at the date $t$ of each of their funding rounds,
    \item bin these observations according to 6-month periods i.e. a funding round announced 200 days after the creation of a startup is counted as an observation for the $T_1=[6,12]$ months post-creation period, while another announced 420 days after creation is counted as an observation for the $T_2=[12, 18]$ months post-creation)
    \item compute the median speed $V_{e}(T)$ for each of these 6-month temporal bins.
\end{itemize} 

\begin{figure}
    \centering
    \includegraphics[scale = .7]{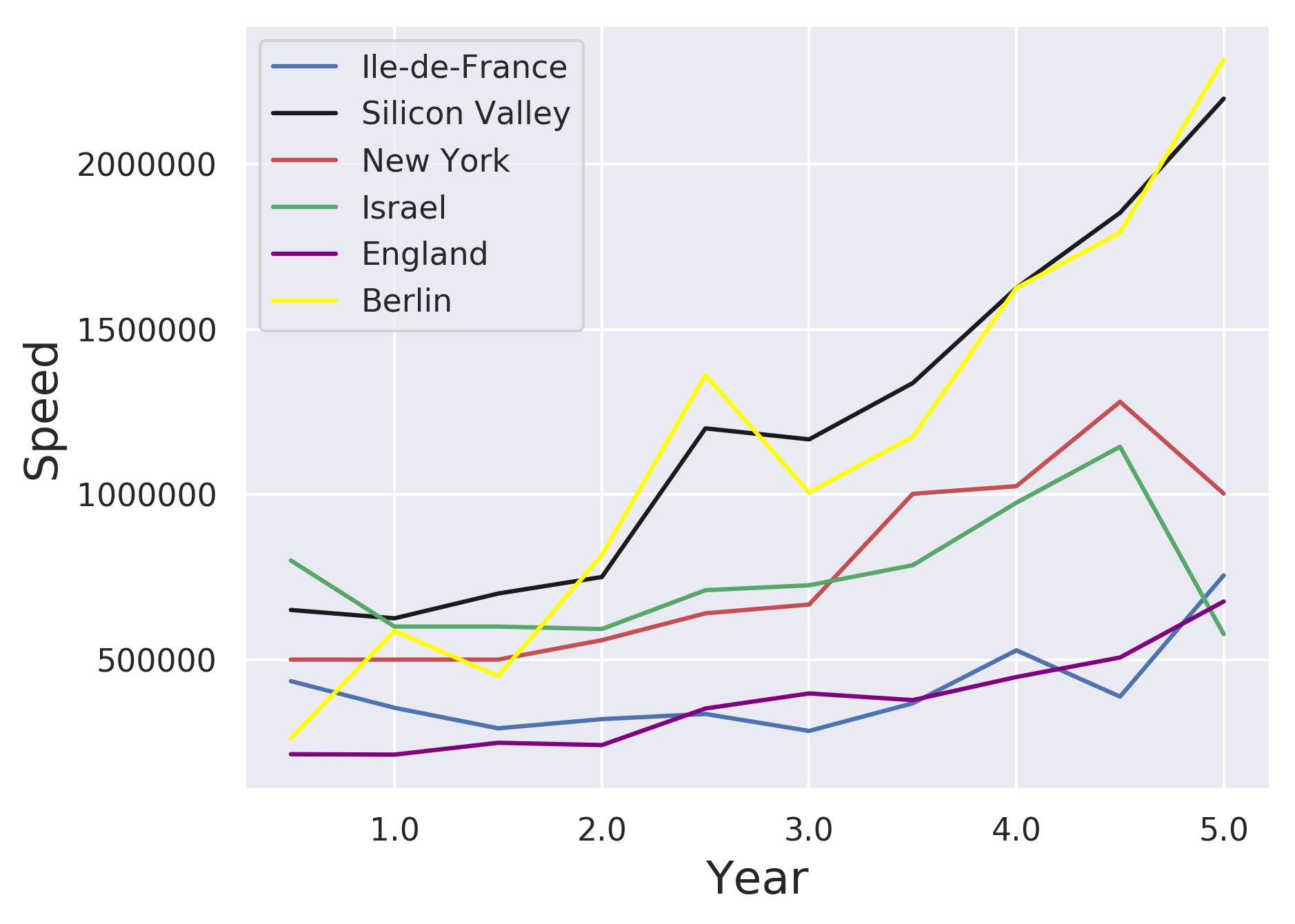}
    \caption{Median fundraising speed $V_{e}(T)$ for 6 entrepreneurial ecosystems (see main text) for $T \in [0,5 years]$ where $T=0$ corresponds to startup creation, for all startups founded in 2010 or after.}
    \label{fig:speed}
\end{figure}

Fig.~\ref{fig:speed} suggests 3 different groups : startups in Berlin and the Silicon Valley appear to receive funding at the fastest speed, while startups in New York and Israel appear to raise funds at a slower pace, and startups in London and Paris at an even slower one.

\subsection{\textit{Acceleration}}

\begin{figure}
    \centering
    \includegraphics[scale = .5]{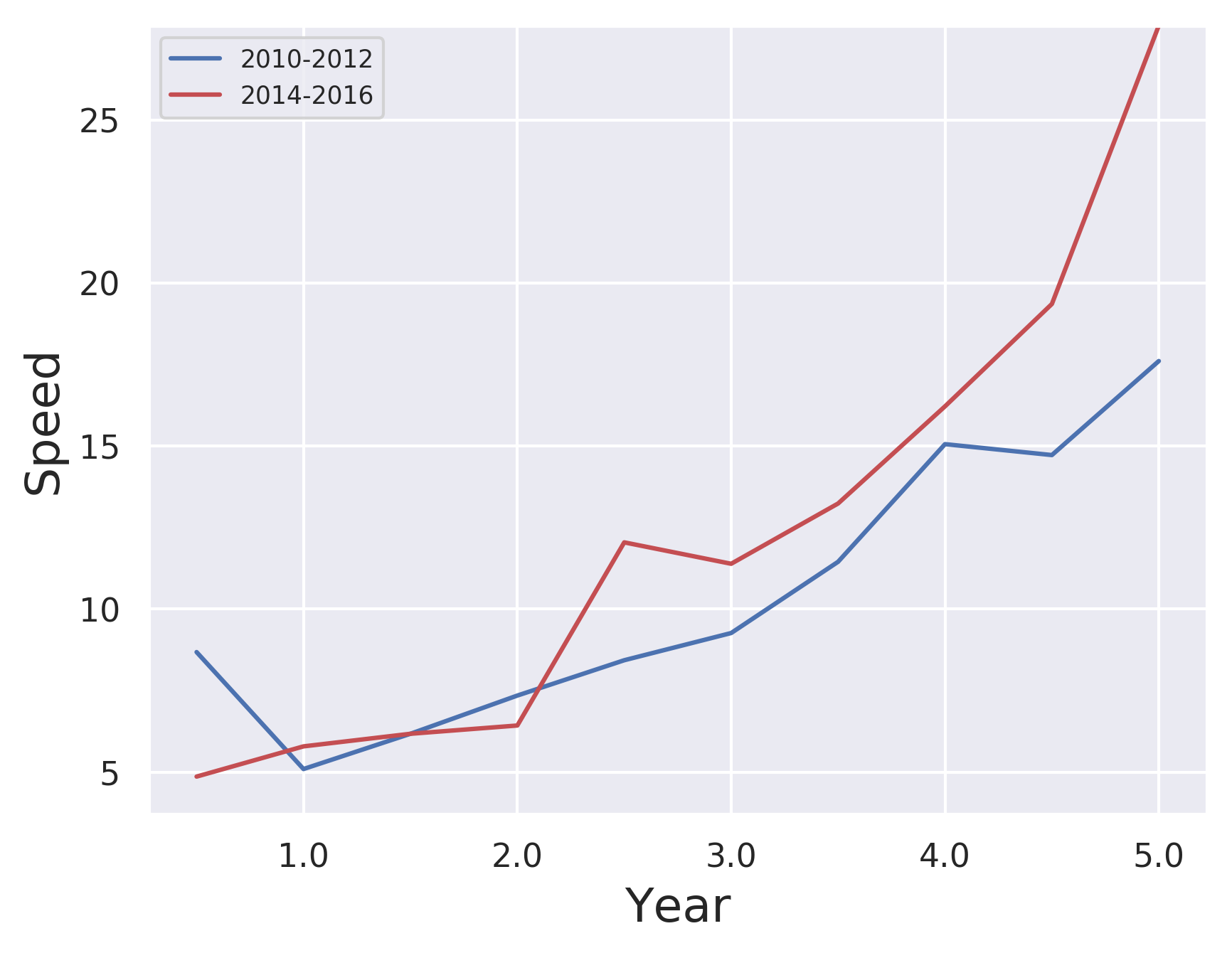}
    \includegraphics[scale = .5]{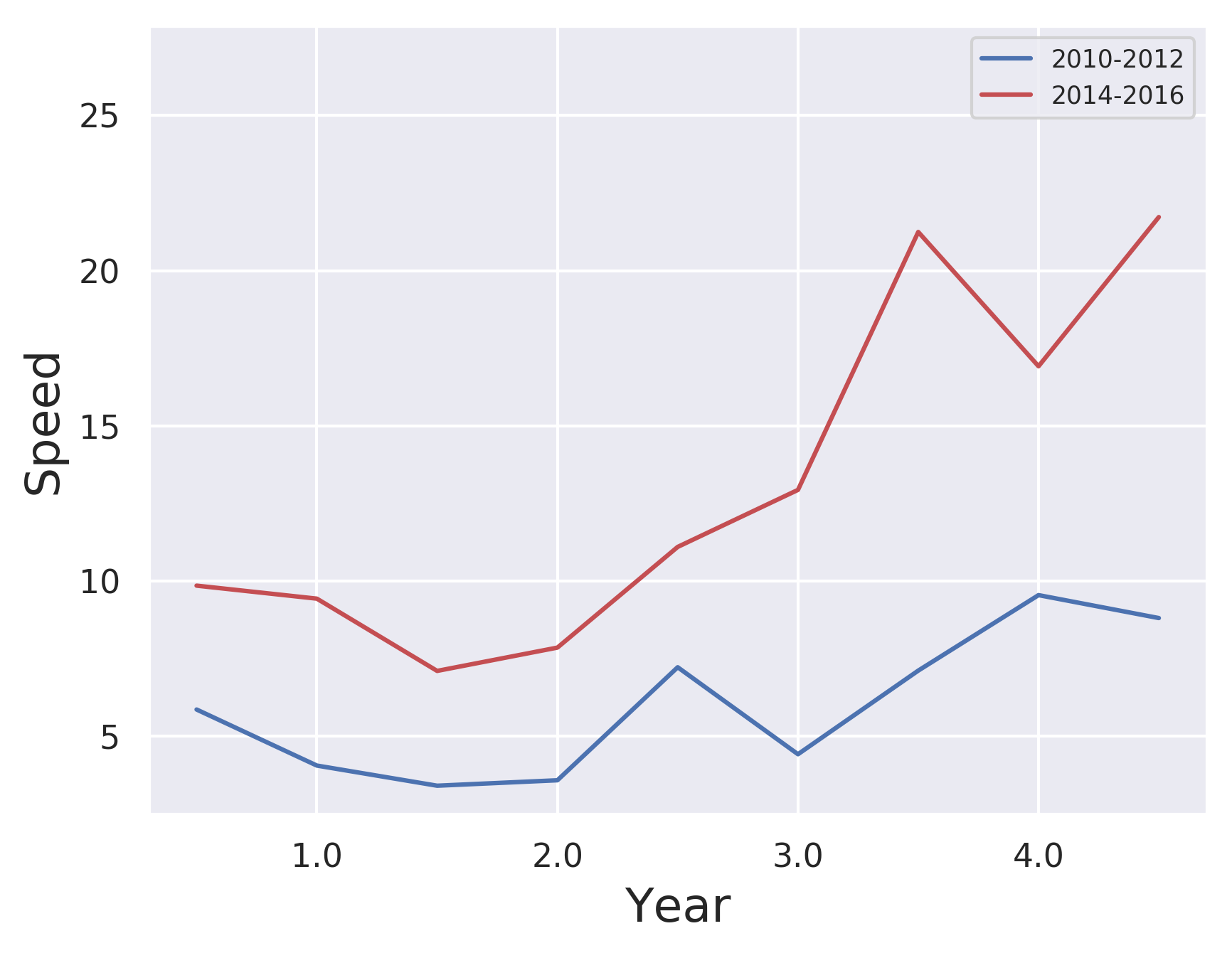}
    \caption{Fundraising speed for the 2010-2012 and the 2014-2016 cohorts (see main text) in the Silicon Valley (left) and in Paris (right). x-axis shows time since startup creation.}
    \label{fig:speeddiff}
\end{figure}

\begin{figure}
    \centering
    \includegraphics[scale = .5]{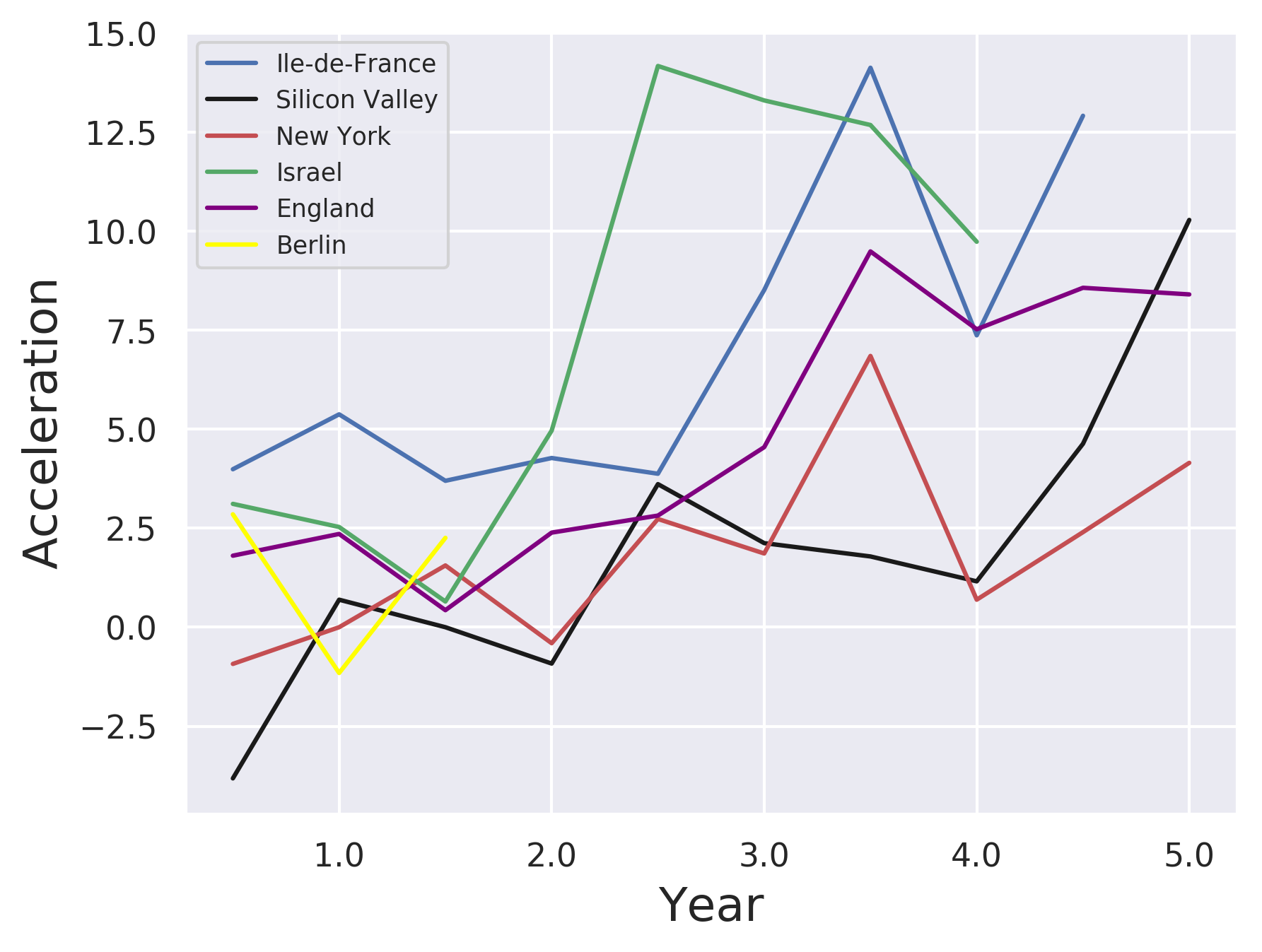}
    \includegraphics[scale = .5]{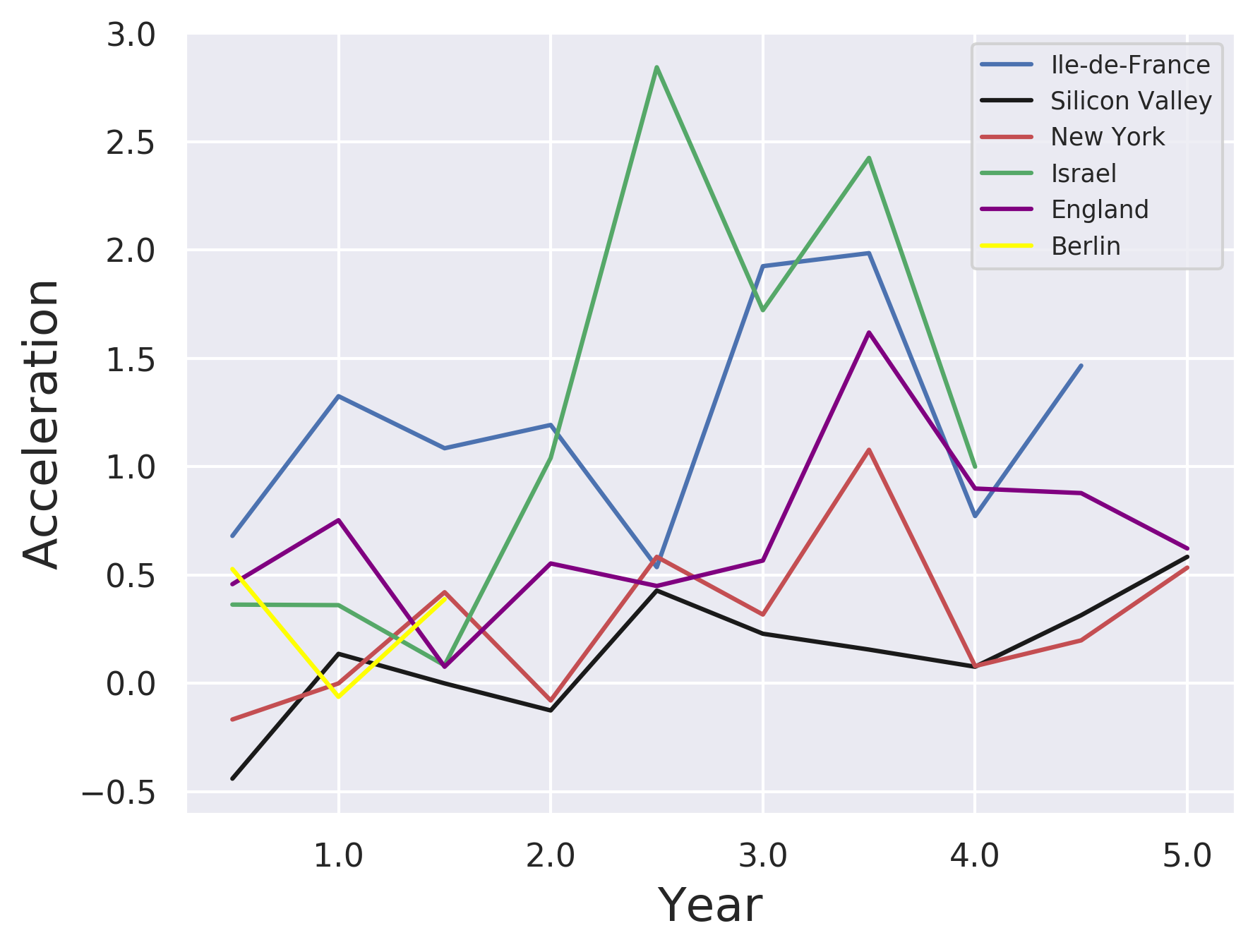}
    \caption{Acceleration for 6 entrepreneurial ecosystems, with respect to the 2010-2012 and the 2014-2016 startup cohorts. x-axis shows time since startup creation. In USD million*year\textsuperscript{-2} (left) and in percentage (right). }
    \label{fig:acceleration}
\end{figure}

Looking at the aggregate speed since 2010 potentially masks a temporal evolution i.e. if \textit{Fundraising Speed} has varied over time. We thus separate two cohorts of startups created at the beginning of our observation period -- between 2010 and 2012, included -- and a few years later -- between 2014 and 2016, included. Fig.~\ref{fig:speeddiff} compares \textit{Fundraising Speed} for these 2 cohorts for 2 selected ecosystems (Silicon Valley and Paris). Fig.~\ref{fig:acceleration} presents the corresponding measures of \textit{Acceleration} for all 6 ecosystems. Although a global acceleration pattern is visible, there are marked differences among ecosystems:
\begin{itemize}
    \item In the Silicon Valley, acceleration appears limited if not absent in the earliest stages of startup development (first 2 years), but more and more pronounced at a later stage, notably 4 years after startup creation.
    \item New York follows an acceleration pattern similar to Silicon Valley's, though without a strong later-stage acceleration.
    \item Paris exhibits a notable acceleration pattern, at both early and later stages.
    \item A similar pattern applies to Israel, though Israel seems to have accelerated less than Paris at the earliest stages after startup creation, but more significantly around 2.5 years after startup creation, before both curves become similar around 3.5 years.
    \item London has accelerated across all years after startup creation, more for later years, comparatively more than the US ecosystems (Silicon Valley, New York), but less than Paris or Israel.
    \item Too few data points unfortunately do not allow for a proper analysis of the acceleration of the Berlin ecosystem.
\end{itemize}



\subsection{\textit{n\textsuperscript{th}-year speed}}
\begin{figure}
    \centering
    \includegraphics[scale = .5]{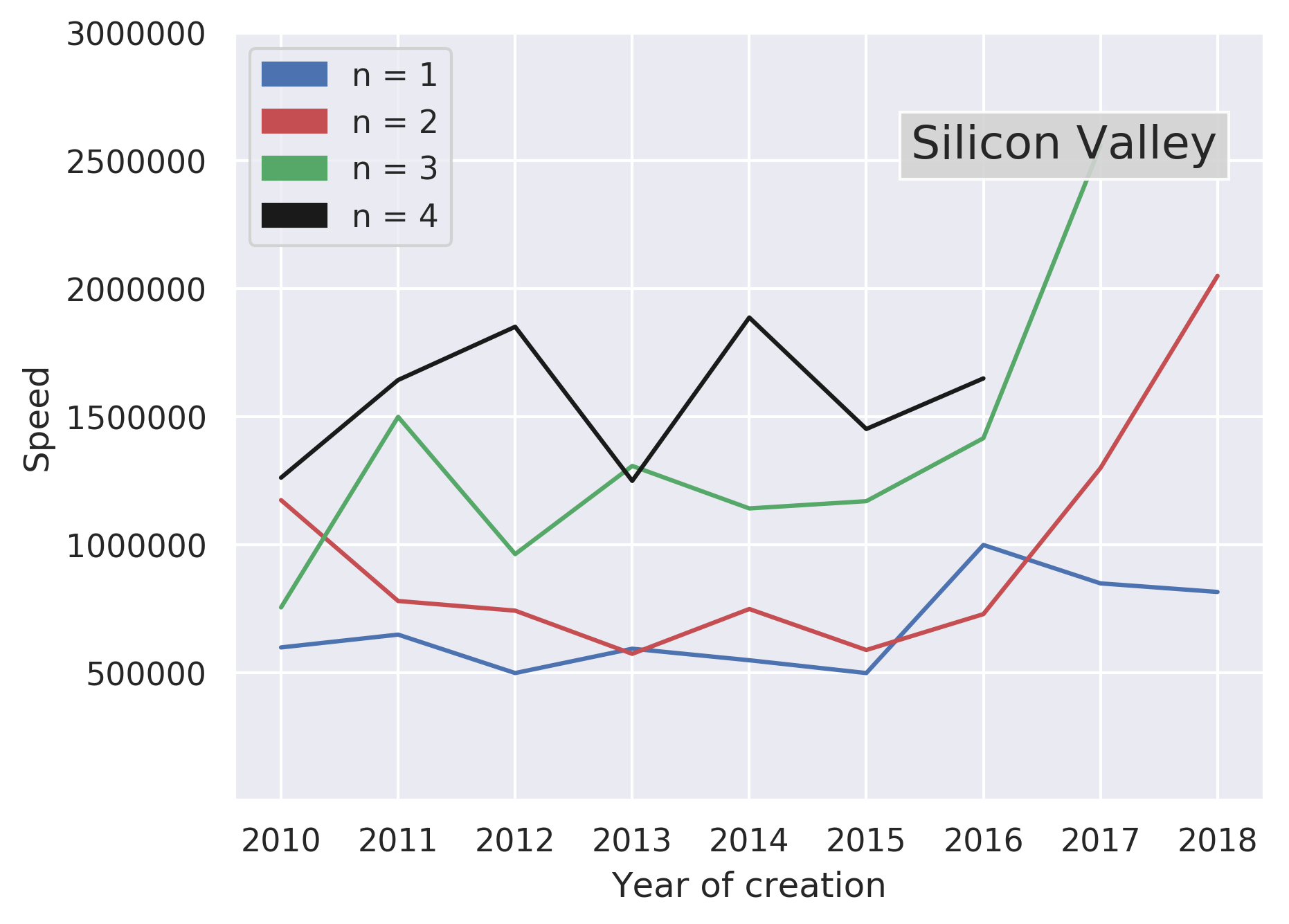}
    \includegraphics[scale = .5]{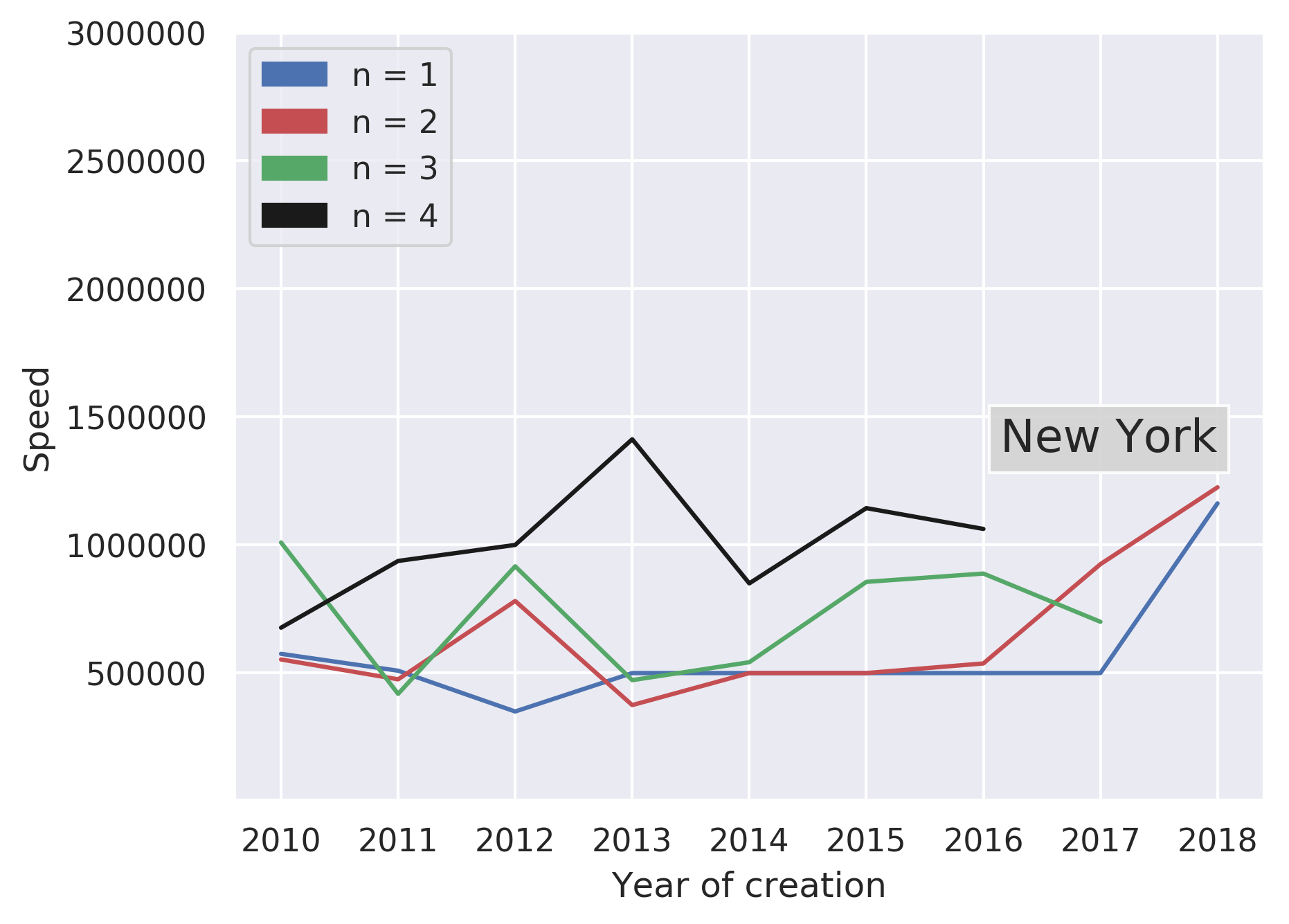}
    \includegraphics[scale = .5]{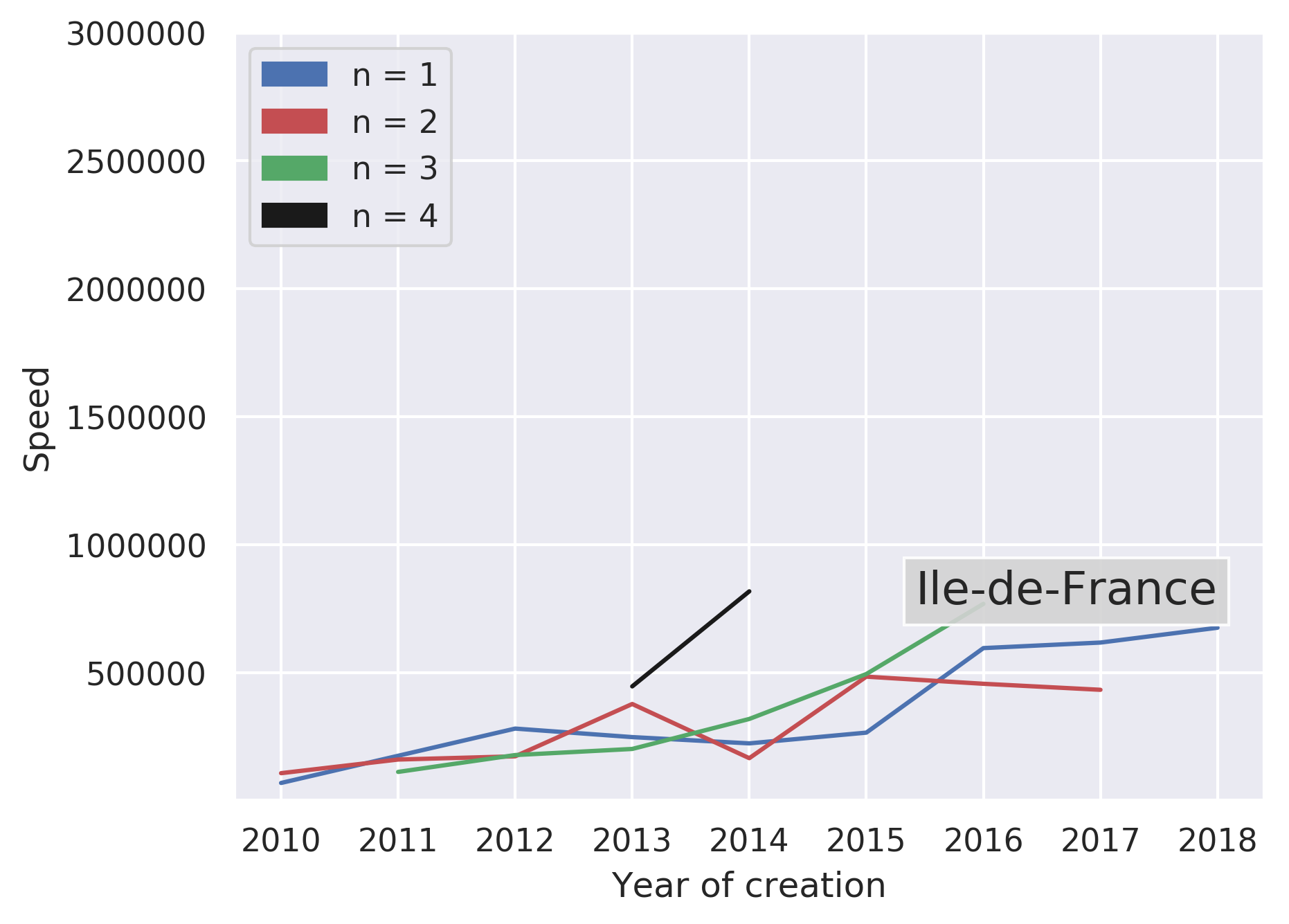}
    \includegraphics[scale = .5]{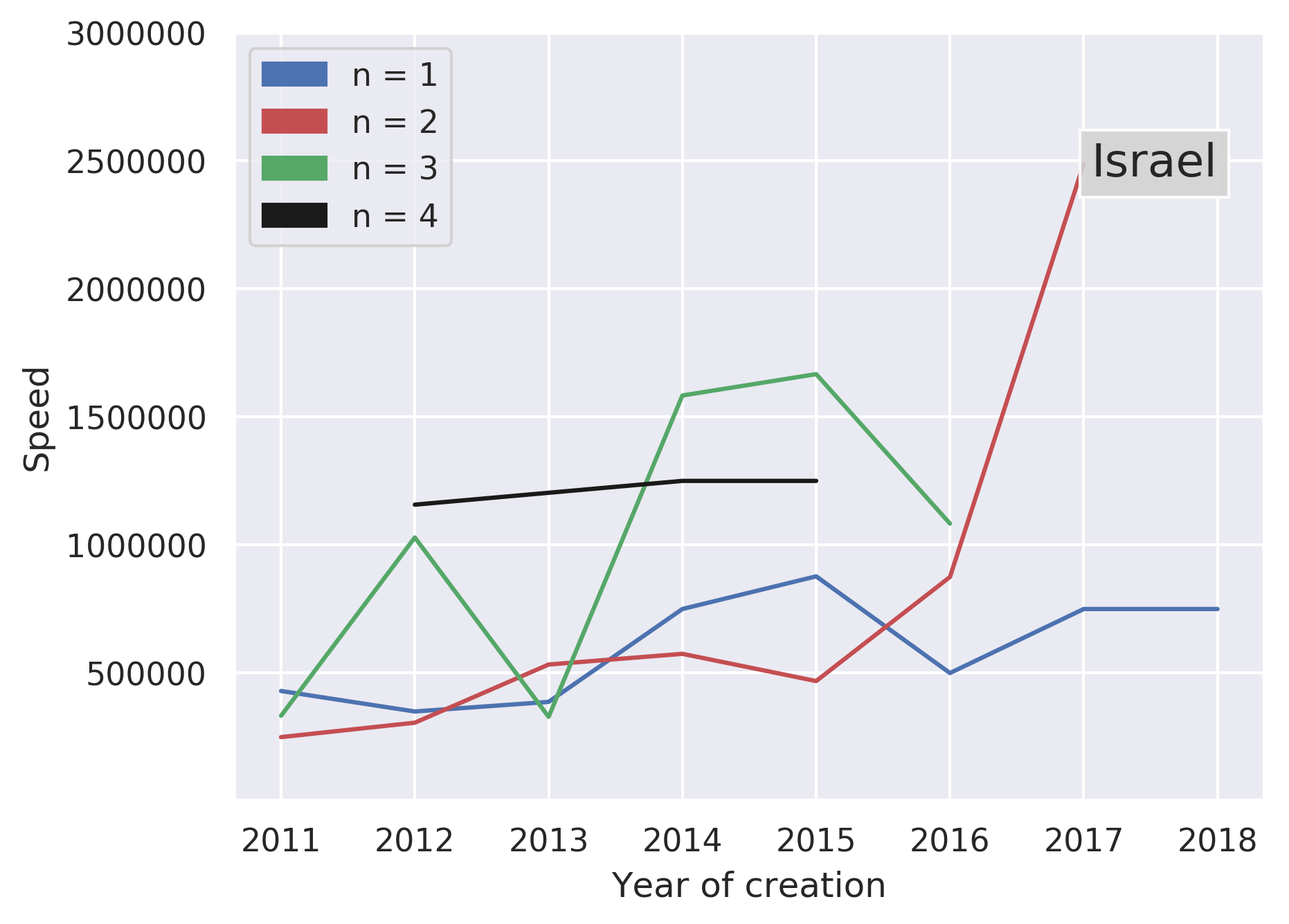}
    \includegraphics[scale = .5]{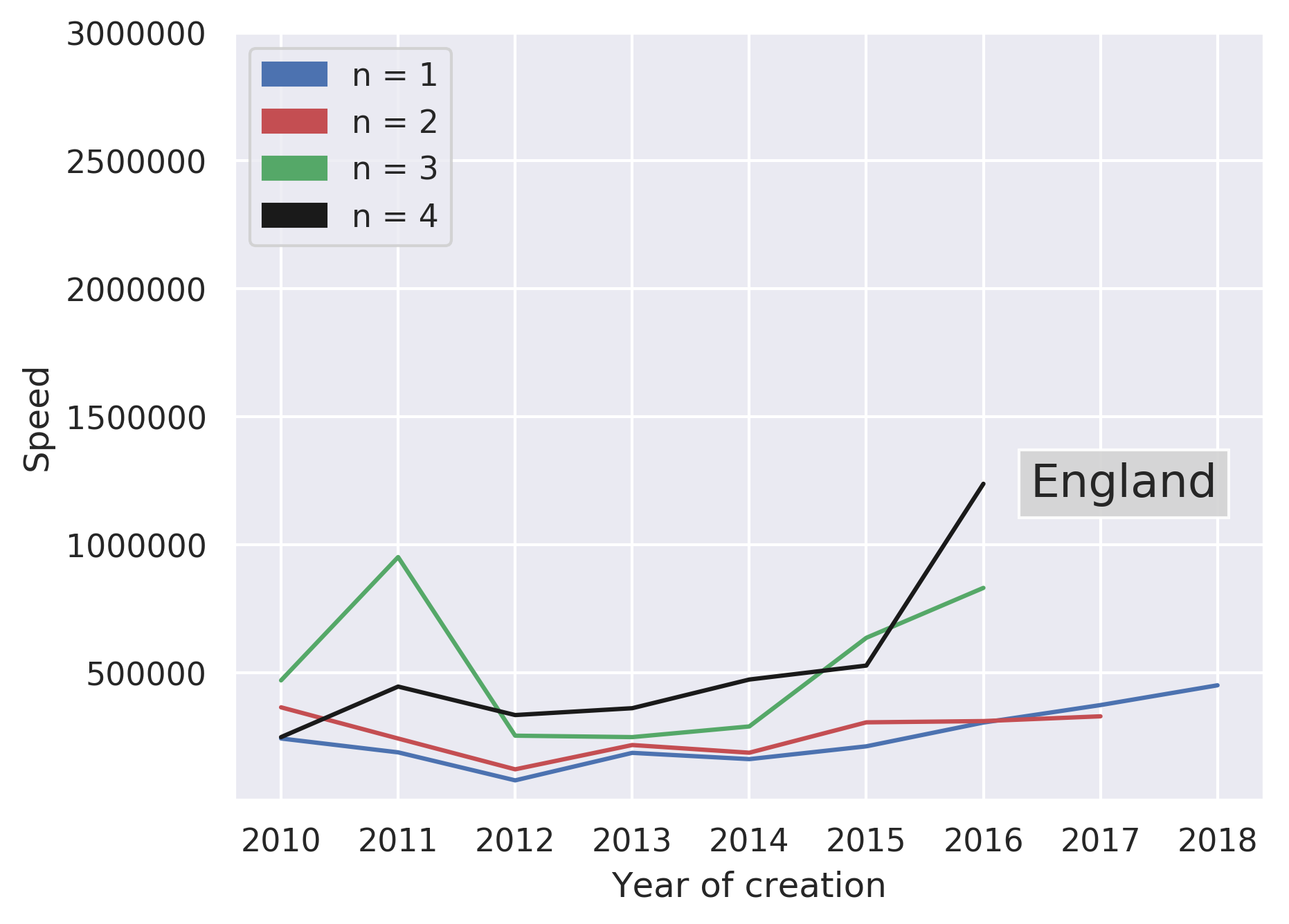}
    \includegraphics[scale = .5]{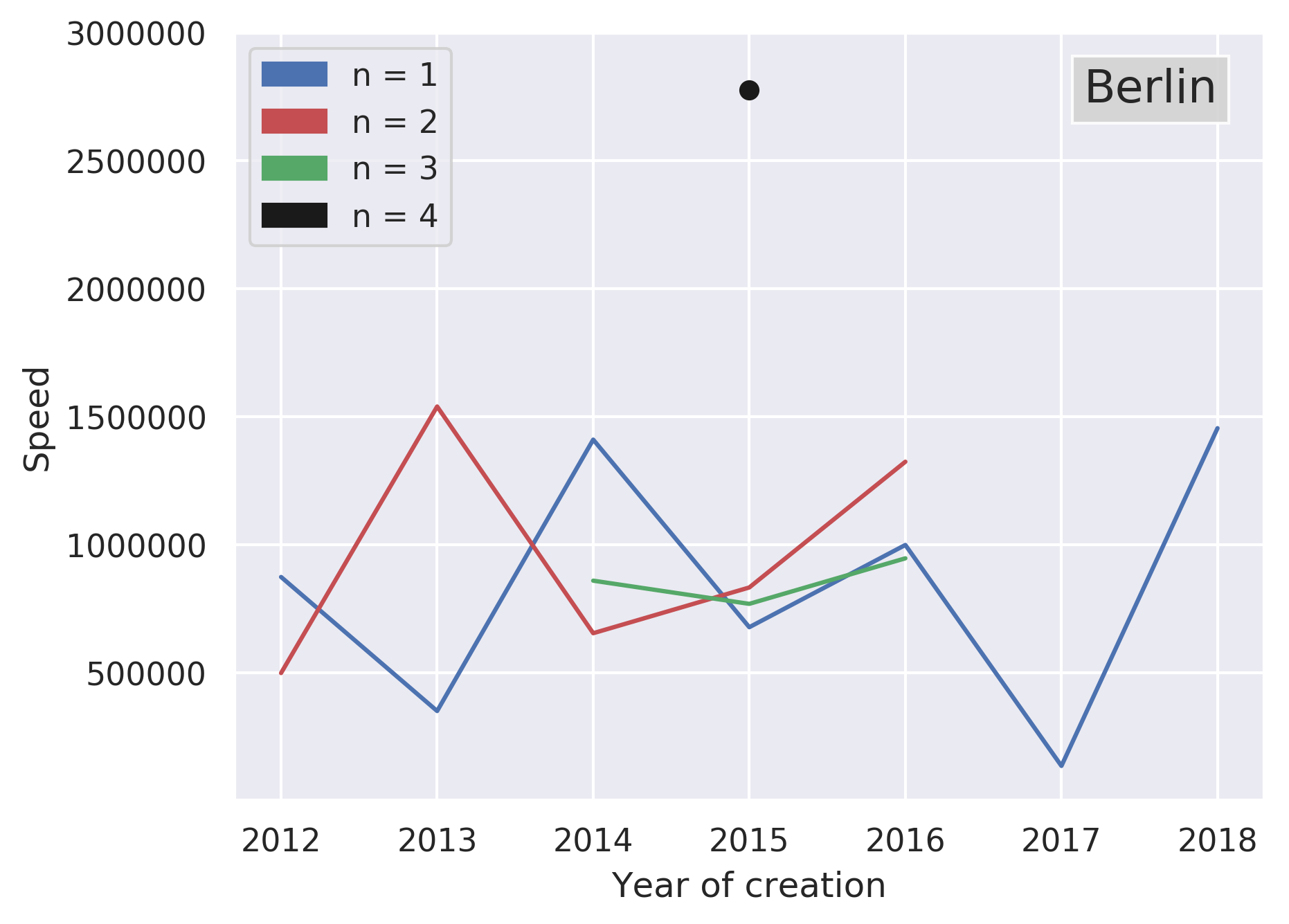}
    \caption{\ $n^{th}-Year Speed$ for 6 entrepreneurial ecosystems (all startups created between 2010 and 2018, included).}
    \label{fig:n_All}
\end{figure}

We next plot \textit{n\textsuperscript{th}-year speed} for each of the 6 studied ecosystems (fig.~\ref{fig:n_All}). Through these lenses, no global acceleration pattern is apparent, except for a mild acceleration in Paris, but with a low baseline (around 100 or 200K USD per year) that is markedly inferior to all other ecosystems, save perhaps for London.

Some ecosystems have seen a marked increase in very recent years, though not enough data is available to confirm this observation: a sharp increase in the Silicon Valley for \textit{2\textsuperscript{nd}-} and \textit{3\textsuperscript{rd}-year speeds}, a more moderate increase in New York for \textit{1\textsuperscript{rst}-} and \textit{2\textsuperscript{nd}-year speeds}, a very sharp increase in Israel for \textit{2\textsuperscript{nd} year speed} and perhaps also for \textit{ 3\textsuperscript{rd}-year speed}, and possibly a moderate increase in London for \textit{4\textsuperscript{th}-year speed}.

In addition, Silicon Valley presents a notable difference between \textit{1\textsuperscript{rst}-year speed} and \textit{2\textsuperscript{nd}-year speed} on the one hand, and \textit{3\textsuperscript{rd}-year speed} and \textit{4\textsuperscript{th}-year speed} on the other, the latter corresponding to a significantly superior fundraising speed. Though less pronounced, this pattern is also visible in London, in New York and perhaps in Paris and in Israel, but only for \textit{4\textsuperscript{th}-year speed}, still taking into account the limitations imposed by our dataset for recent years and for smaller ecosystems in terms of number of startups.

\subsection{'Purchasing Power Parity' and Distribution of investments across funding rounds}
\label{subsec:specific_results}
However, the results presented above do not account for two additional and potentially relevant factors:
\begin{itemize}
    \item 'purchasing power parity'~\cite{Vachris} for the different ecosystems: the cost of living in the Silicon Valley is notoriously high~\cite{Adrian}, compared to Berlin for instance. Comparing raw amounts in different ecosystems can thus be misleading in terms of the comparative opportunities a given amount offers to startups.
    \item the distribution of investments among the various funding stages of the venture capital cycle. A different frequency of different types of rounds across entrepreneurial ecosystems might indeed result in different median speeds in the local population of startups for pure statistical reasons.
\end{itemize}

In order to account for a different 'purchasing' power in terms of raised capital across entrepreneurial ecosystems, we divide raw funding amounts by the average price of a software engineer in each of the ecosystems, considered here as a proxy for 'purchasing power parity', as taken from \hyperlink{https://www.daxx.com/blog/development-trends/it-salaries-software-developer-trends-2019}{daxx.com}, and considered constant for each ecosystem during the period studied.

\begin{figure}[H]
    \centering
    \includegraphics[scale = .7]{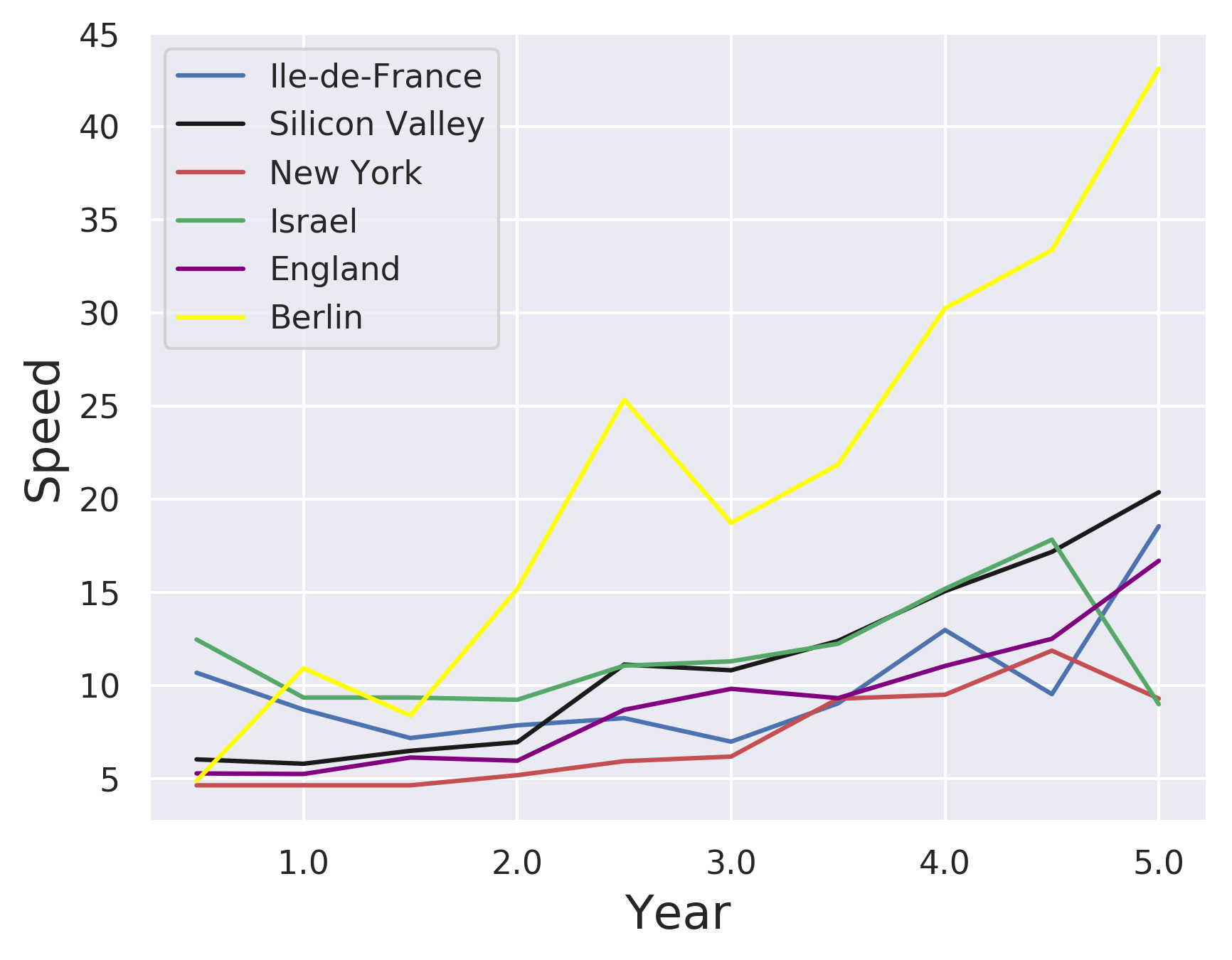}
    \caption{Median fundraising speed $V_{e}(T)$ for 6 entrepreneurial ecosystems for $T \in [0,5 years]$ where $T=0$ corresponds to startup creation, for all startups founded in 2010 or after, taking into account 'purchasing power parity' based on the local cost of a software engineer (see main text).}
    \label{fig:speedppp}
\end{figure}

As seen in fig.~\ref{fig:speedppp}, Berlin is the fastest entrepreneurial ecosystem when the differential cost of living is accounted for, followed by the Silicon Valley and Israel, and then by London, Paris and New York.

\begin{figure}[H]
    \centering
    \includegraphics[scale = .35]{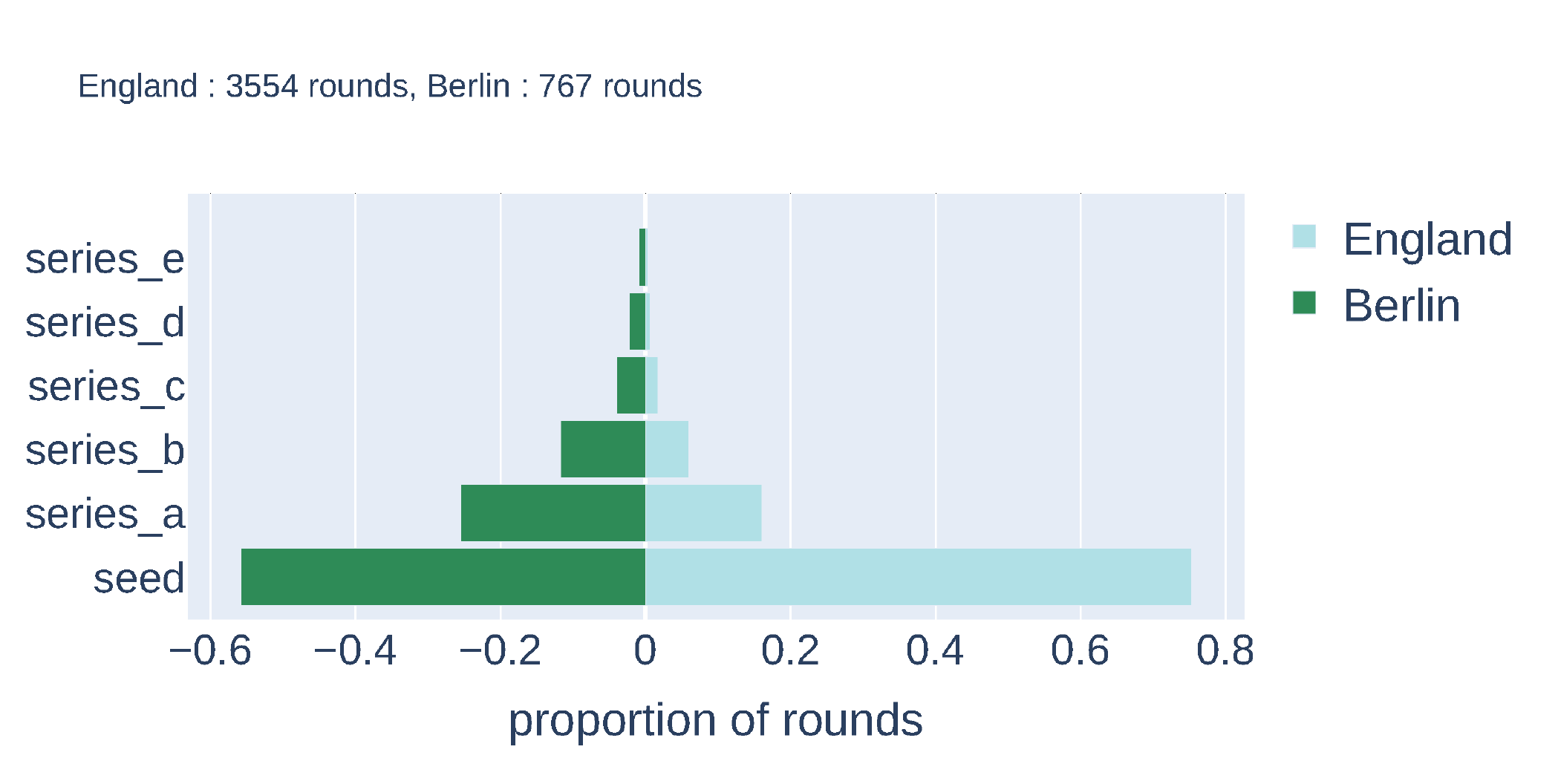}
    \includegraphics[scale = .35]{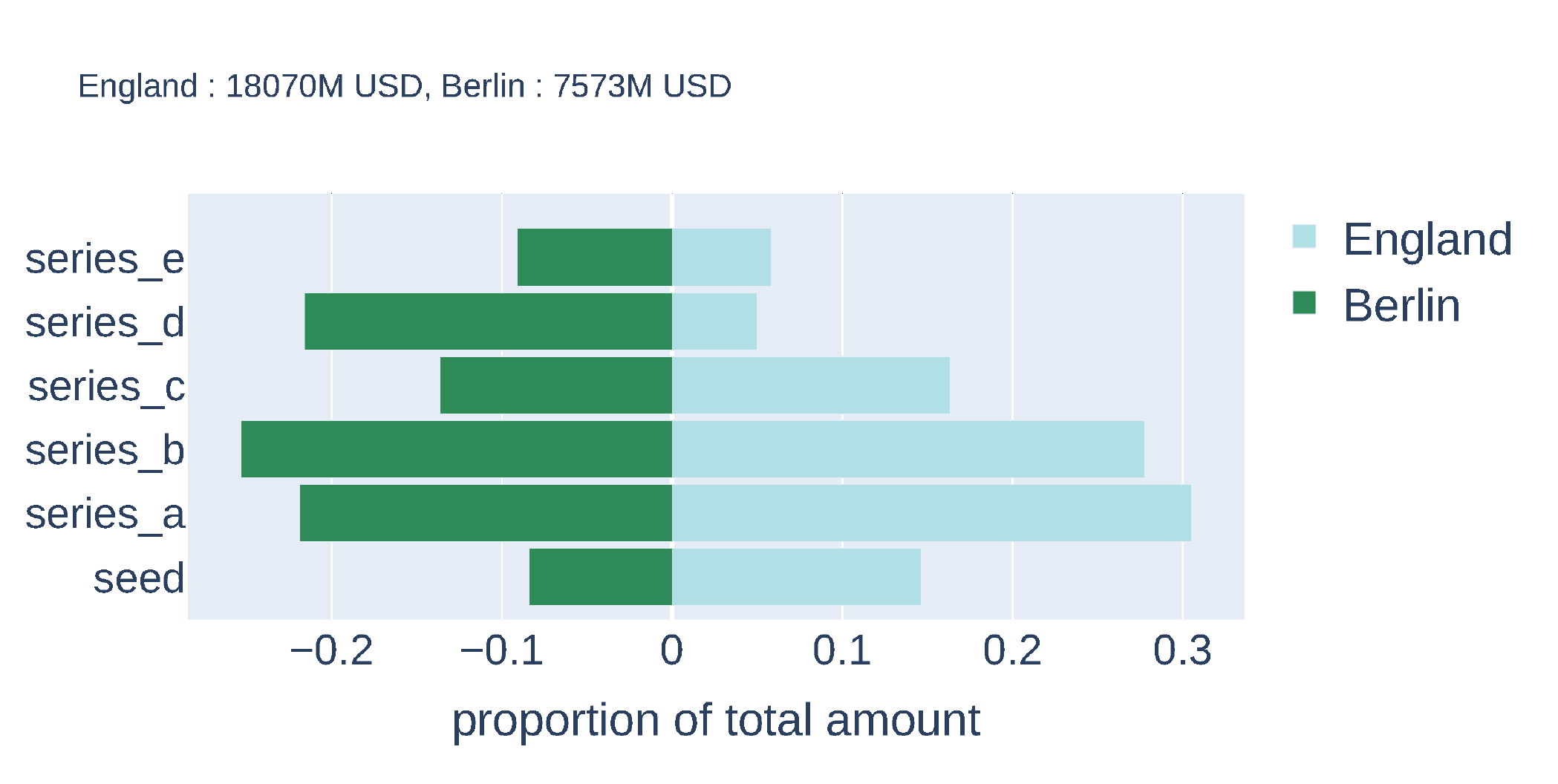}
    \includegraphics[scale = .35]{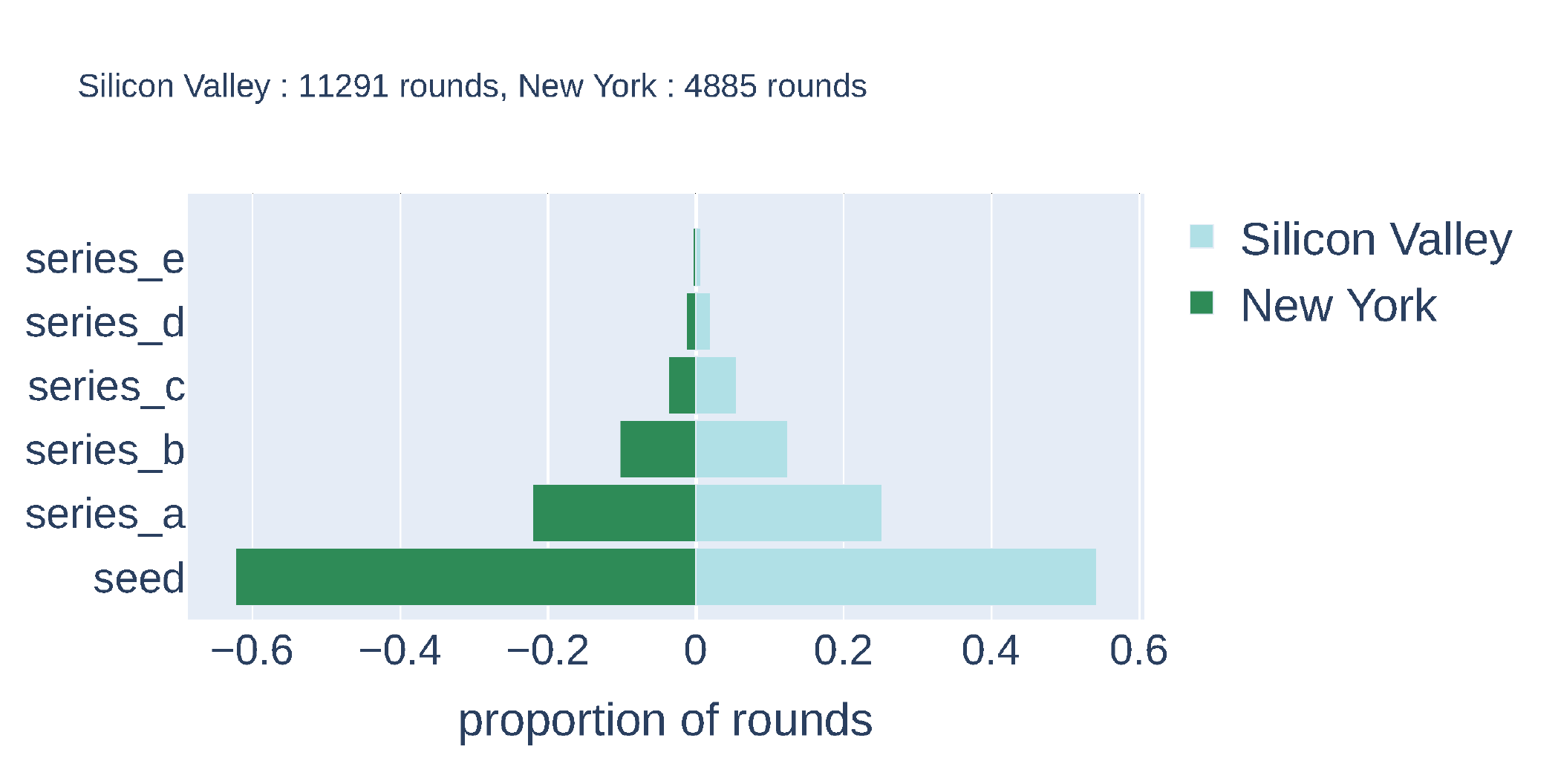}
    \includegraphics[scale = .35]{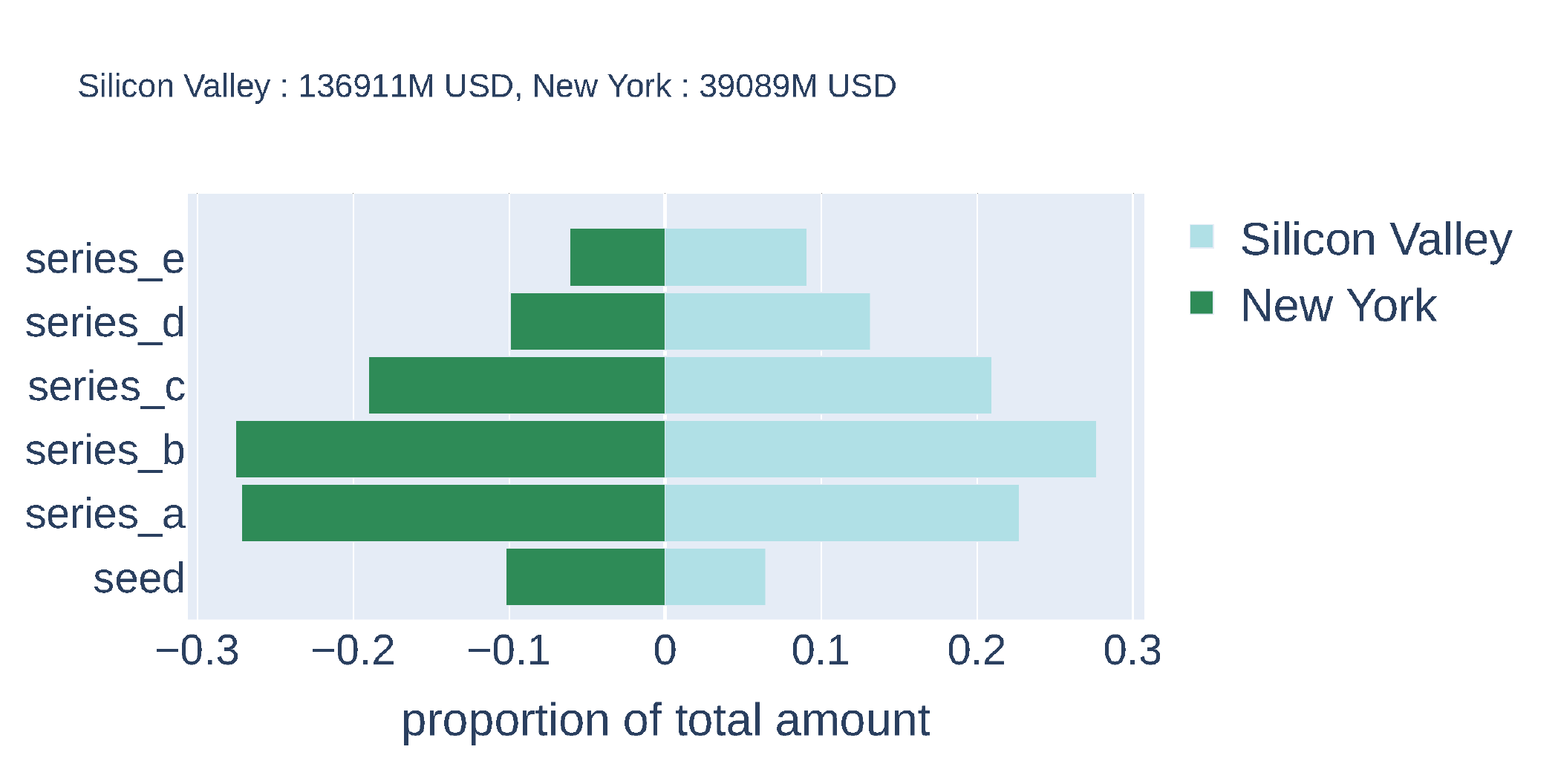}
    \includegraphics[scale = .35]{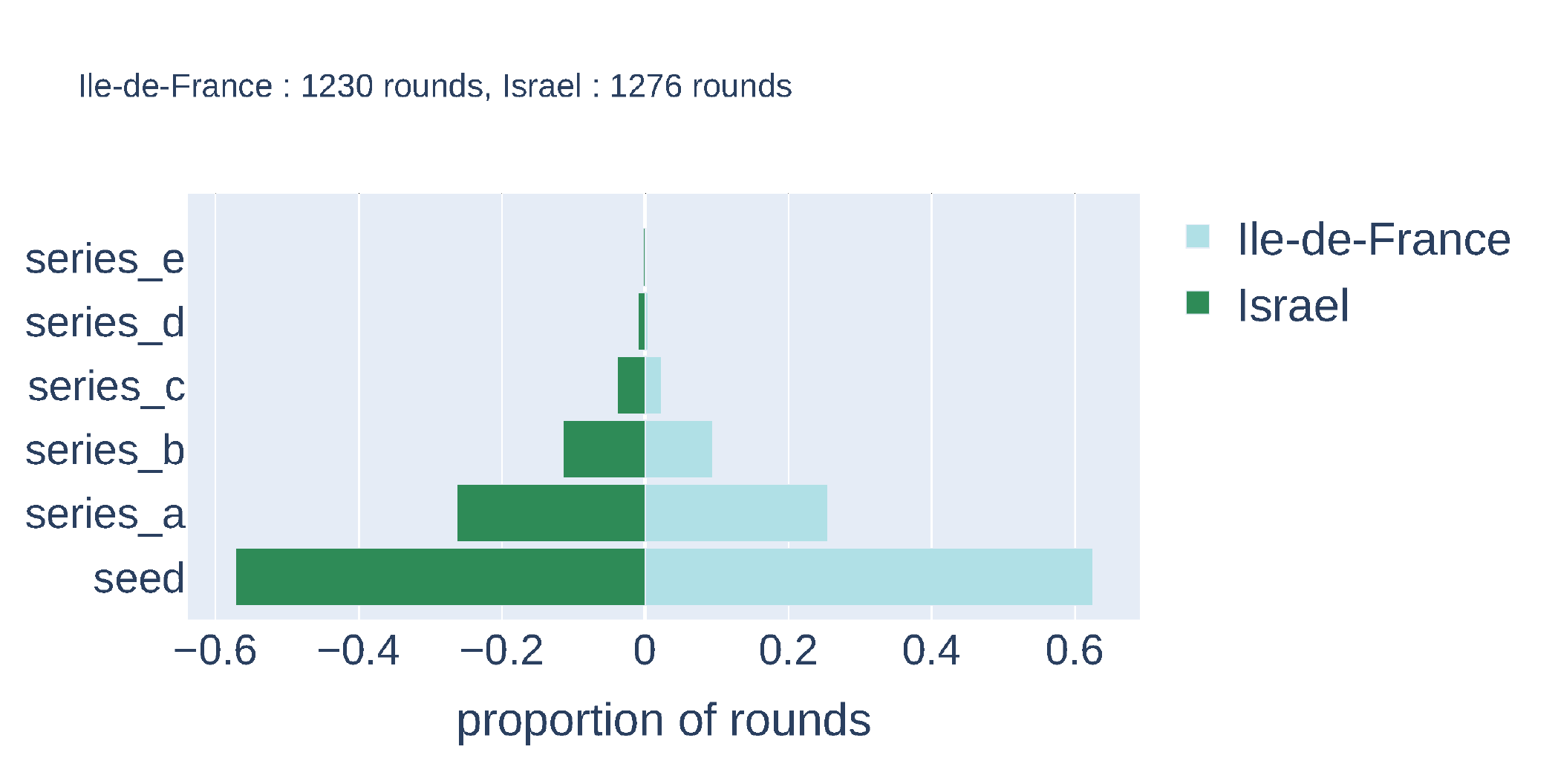}
    \includegraphics[scale = .35]{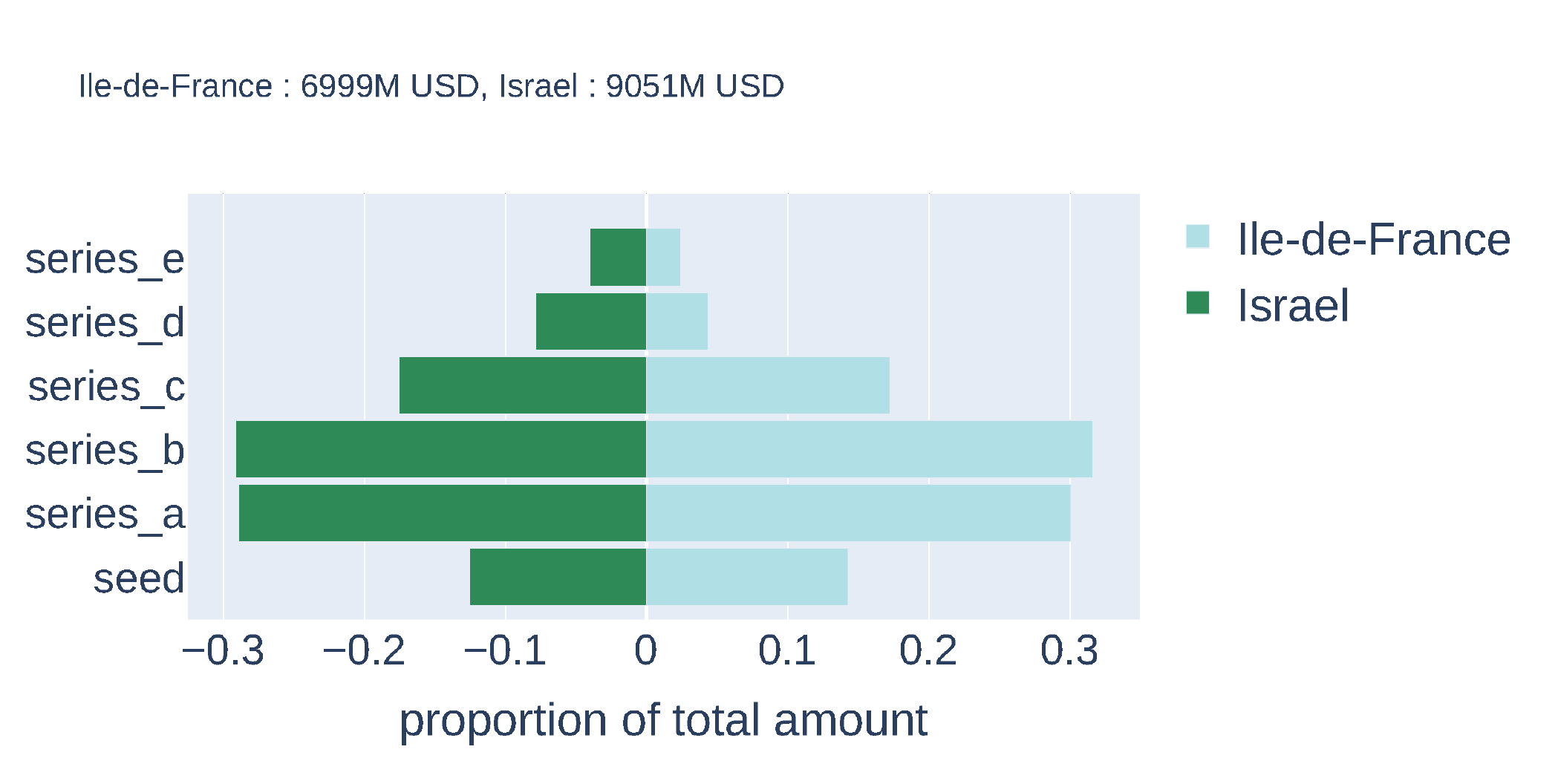}
    \caption{Distribution of the proportion of total funding amount (top) and of number of funding rounds (bottom) for the Berlin ecosystem (green, left) and the London ecosystem (cyan, right) on the 2010-2020 period.}
    \label{fig:distribution}
\end{figure}

Looking now at the distribution of funding amounts and number of funding rounds across entrepreneurial ecosystems, fig. ~\ref{fig:distribution} presents pairwise comparisons of the frequencies of funding rounds (left) and of the relative amounts invested at different rounds (right), presented here in a way similar to age pyramids, of London vs. Berlin, Silicon Valley vs. New York and Paris vs. Israel, respectively. London stands out as the entrepreneurial ecosystem where seed rounds are the most frequent compared to others, while a significant share of total investment is allocated to seed stage, again compared to other ecosystems. Berlin and the Silicon Valley stand out as concentrating the most important share of their total funding on the later rounds (D and E), while the share of the total amount invested in the Silicon Valley at seed stage is comparatively the lowest and the share of the total amount invested in Paris at later stages (D and E) is also the lowest, compared again to other ecosystems.


\section{Interpretation and Conclusion}
Altogether, the results presented above that have tried to characterize 6 prominent entrepreneurial ecosystems using indicators suggest very different acceleration strategies across ecosystems:
\begin{itemize}
    \item In Berlin, a concentration of investment at the later stage associated with fewer startups funded results in a very high fundraising speed, even more so when taking 'purchasing power parity' into account. This pattern does not seem to have changed over the past decade: no acceleration is visible.
    \item In the Silicon Valley, a high fundraising speed is somewhat counterbalanced by a high standard of living. Startups reach very high level of fundraising speed when they are only 3 years old, and it still has accelerated considerably in recent years. Seed stage comparatively represents a lower share of all investments compared to other ecosystems, while later stage funding is comparatively favored.
    \item In London, numerous startups receive seed funding, which represents a relatively important share of the total amount invested in this ecosystems. As a result, their median speed is low at the earlier stage, but startups reach higher speed levels as soon as they are 3 years old, and a global acceleration pattern is visible over time.
    \item In New York, startups have to wait until their 4\textsuperscript{th} year to reach higher speed levels. No global acceleration pattern is visible, except perhaps in the most recent years.
    \item In Israël, a sharp acceleration over the past decade is visible for 2\textsuperscript{nd} and 3\textsuperscript{rd} year startups. The global fundraising speed makes it one of the fastest ecosystems, on par with the Silicon Valley when taking 'purchasing power parity' into account.
    \item In Paris, finally, a particularly low fundraising speed is not counterbalanced by a favorable 'purchasing power parity' nor by a clear acceleration pattern over the past decade, quite remarkable among studied ecosystems but whose limit stems not only from the initial conditions in Paris -- a very low fundraising speed -- but also from an allocation of funding amounts that favors seed stage and that tends to discriminate against later stages. Altogether, Parisian startups do not benefit from an increased fundraising speed until late, at best in their 4\textsuperscript{th} year.
\end{itemize}

Obviously, these observations easily resonate with several stylized facts gathered from more anecdotal evidence and very often echoed within entrepreneurial ecosystems. For instance, the fact that the Paris ecosystem could succeed in creating local champions but would generally fail to create global champions, as it lacks the financial means to create global leaders: Parisian startups are not accelerated enough, compared to their competitors from other ecosystems, which might explain why so many of the recent unicorns in France have raised money abroad. Or else, the ability of the Silicon Valley to allow its startups to reach high speed levels very early on, that could reflect a culture more prone to taking risks among its local venture capital community -- then with a similar and perhaps related pattern in Israel. Conversely, the results associated with the Berlin ecosystem in terms of its ability to create fast-growing startups are probably more surprising and reflect an acceleration strategy aligned with the general capability of the German economy to create medium-sized companies known as the Mittelstand and to help them thrive -- here, a capability that appears to have started to also benefit startups.

However, several other investigations are obviously needed before more definite conclusions can be reached, and other indicators might be helpful in this respect. Our aim here was not to reach a definite conclusion, but rather to show that simple yet powerful indicators could play a very significant role on the way to such analyses of the relative performance of entrepreneurial ecosystems, and to also suggest that different entrepreneurial ecosystems can be associated with different acceleration strategies. Such a finding could help policymakers take such additional evidence in consideration when designing newer steps for local, national and federal innovation policies.

\end{document}